\documentclass[prd,twocolumn,superscriptaddress,floatfix,amsmath,amssymb,amsfonts,longbibliography,nofootinbib]{revtex4-2}

\usepackage{float} 

\usepackage[normalem]{ulem}
\usepackage{graphicx}
\usepackage{dcolumn}
\usepackage{bm}
\usepackage{blindtext}
\usepackage{verbatim}
\usepackage{relsize}
\usepackage{mathrsfs}
\usepackage{musicography}
\usepackage{amsmath}
\usepackage{blindtext}
\usepackage{cancel}
\usepackage{physics}
\usepackage{epstopdf}
\usepackage{mathtools}
\usepackage{tensor}
\usepackage{color}
\usepackage[usenames,dvipsnames]{pstricks}
\usepackage{epsfig}
\usepackage{pst-grad} 
\usepackage{pst-plot} 
\usepackage{hyperref}
\hypersetup{
    colorlinks=true,
    linkcolor=blue,
    filecolor=magenta,      
    urlcolor=cyan,
}
\usepackage{verbatim}
\usepackage{slashed}

\usepackage{dsfont}

\newcommand{\ts}{\textsc}
\newcommand{\hatr}{\hat{r}}

\newcommand{\rhoh}{\hat{\rho}}
\newcommand{\muh}{\hat{\mu}}
\newcommand{\mf}{\mathsf}

\newcommand{\ii}{\mathrm{i}}

\renewcommand{\L}{\mathcal{L}}

\usepackage{xcolor}

\newcommand{\proj}[2]{\ket{#1} \! \bra{#2}}

\begin{document}

\title{Entanglement harvesting: state dependence and covariance}

\author{H\'ector Maeso-Garc\'ia}
\email{hmaesoga@uwaterloo.ca}

\affiliation{Centre de Formació Interdisciplinària Superior, Universitat Politècnica de Catalunya, 08028 Barcelona, Spain}
\affiliation{Department of Applied Mathematics, University of Waterloo, Waterloo, Ontario, N2L 3G1, Canada}

\author{Jos\'e Polo-G\'omez} 
\email{jpologomez@uwaterloo.ca}

\affiliation{Department of Applied Mathematics, University of Waterloo, Waterloo, Ontario, N2L 3G1, Canada}
\affiliation{Institute for Quantum Computing, University of Waterloo, Waterloo, Ontario, N2L 3G1, Canada}
\affiliation{Perimeter Institute for Theoretical Physics, Waterloo, Ontario, N2L 2Y5, Canada}

\author{Eduardo Mart\'{i}n-Mart\'{i}nez}
\email{emartinmartinez@uwaterloo.ca}

\affiliation{Department of Applied Mathematics, University of Waterloo, Waterloo, Ontario, N2L 3G1, Canada}
\affiliation{Institute for Quantum Computing, University of Waterloo, Waterloo, Ontario, N2L 3G1, Canada}
\affiliation{Perimeter Institute for Theoretical Physics, Waterloo, Ontario, N2L 2Y5, Canada}

\begin{abstract}
    We analyze entanglement harvesting for arbitrary  initial states of particle detectors and arbitrary quasifree states of the field. Despite the fact that spatially smeared particle detectors are known to break covariance for arbitrary initial states, we show that the entanglement harvested by a pair of detectors from a quasifree state of a scalar field is a covariant quantity up to second order in perturbation theory. 
\end{abstract}

\maketitle

\section{Introduction}

Quantum fields are generically entangled across different regions of spacetime~\cite{SUMMERS, vacuumEntanglement,Bombelli1986,witten}. However, quantifying this shared entanglement is a hard problem that remains unsolved for arbitrary pairs of regions~\cite{witten}. Even for the simpler case of complementary regions, it has only been successfully evaluated in particular cases (e.g.,~\cite{Calabrese2004,Ryu2006,Calabrese2009,Saravani2014}).

In this context, particle detector models, such as the Unruh-DeWitt (UDW) model~\cite{Unruh1976,DeWitt}, can provide an efficient means of accessing this entanglement structure. Specifically, a pair of particle detectors, i.e., localized non-relativistic quantum systems, can couple to a quantum field extracting a certain amount of entanglement from it~\cite{Valentini1991,Reznik2003,Reznik1,Nick,Pozas-Kerstjens:2015}. When they are spacelike separated, the entanglement that the detectors acquire is coming from pre-existing entanglement in the field between the spacetime regions to which the detectors coupled. This phenomenon is known in relativistic quantum information as \textit{entanglement harvesting}, and it has been profusely studied both in itself (see, among many others,~\cite{Reznik2,Edu2012,Salton:2014jaa,Pozas2016,PetarOld,Petar,Henderson2018, Ng2018,Ng2,Henderson2019,Cong2019,Cong2020,Henderson2020,ericksonNew,Liu2021,carol}), to use it as a resource~\cite{Edu2013farming,Brown2014} or as a quasi-local witness of, e.g., spacetime geometry~\cite{Nick} and topology~\cite{topology}. 


Since particle detectors are non-relativistic systems one could wonder whether coupling them to quantum fields can cause any problems with causality or general covariance. It has been shown that particle detectors are both free from causality issues~\cite{martin-martinez2015,pipoFTL} and their coupling is generally covariant~\cite{us,us2}---in an exact form when the detectors are pointlike, and in an approximate way when the detectors are smeared. The smearing can sometimes enable a certain degree of faster-than-light signalling (FTL), and lead to non-covariance issues. However, such contributions are suppressed with the lengthscale of the smearing in both cases, and they are often subleading within the usual perturbative approximations~\cite{us2,pipoFTL}.

In particular, setups involving only two (compactly supported) particle detectors have been shown to be safe from faster-than-light signalling~\cite{martin-martinez2015,pipoFTL}, and therefore so is entanglement harvesting. Even if more detectors are considered---within the usual perturbative regimes\mbox{---,} entanglement harvesting is a second order process, while any possible FTL signalling between smeared detectors is at least of fourth order and therefore subleading.

In contrast, the breakdown of covariance found in~\cite{us2} for scenarios with multiple detectors does show up in general at second order in perturbation theory, when the initial states of the detectors do not commute with their free Hamiltonians. Since entanglement harvesting is a second order process, the question remains whether one could use smeared detectors without having to worry that the model may not be covariant. 

In this manuscript we address the question of whether these covariance issues affect our predictions of the entanglement harvested at leading order, showing that this is not the case. As a necessary step to do so, we also generalize previous results on entanglement harvesting considering arbitrary initial uncorrelated pure states for the detectors, as well as general mixed states when the amount of mixedness is small.

\section{Setup} 

For the entanglement harvesting protocol, we consider a real scalar field $\hat{\phi}(\mf x)$ in a $1+d$ dimensional globally hyperbolic spacetime $(M, \mf{g})$, where $\mf {g}$ is a Lorentzian metric. We consider two two-level UDW particle detectors \cite{Unruh1976,DeWitt}, which we denote as $\ts{A}$ and $\ts{B}$, undergoing timelike trajectories $\mf{z}_{\textsc{a}}(\tau_{\textsc{a}})$ and $\mf{z}_{\textsc{b}}(\tau_{\textsc{b}})$ parametrized by their proper times, and with ground states $\ket{g_{\ts{a}}}$ and $\ket{g_{\ts{b}}}$, excited states $\ket{e_{\ts{a}}}$ and $\ket{e_{\ts{b}}}$, and proper energy gaps $\Omega_{\ts a}$ and $\Omega_{\ts b}$, respectively. We consider the UDW detectors to be initially in an arbitrary product state
\begin{equation}
    \rhoh_{\textsc{ab},0} = \rhoh_{\ts{a},0} \otimes \rhoh_{\ts{b},0}.
\end{equation}    
In the literature, these initial states are commonly taken to be the ground  (or excited) eigenstates of the detectors (see, e.g.,~\cite{Reznik2003,Reznik1,Pozas-Kerstjens:2015,Salton:2014jaa,Pozas2016,Nick,PetarOld,Petar, Ng2,Henderson2019,topology,ericksonNew}, among many others). These particular cases were already shown in~\cite{us2} to be safe from covariance violations up to second order.

To analyze more general states, let us begin from the most general scenario where the quantum field and the two detectors start uncorrelated. The initial density operator of the joint particle detectors-field system is
\begin{equation}
    \rhoh_{0} = \rhoh_\textsc{ab,0} \otimes \rhoh_{\phi,0}.
\end{equation}
We will assume that the initial field state $\rhoh_{\phi,0}$ is a quasifree (zero-mean Gaussian) state, i.e., its odd-point functions are zero, and its even-point functions can be written in terms of the two-point function \mbox{$W(\mf x, \mf x') = \text{Tr}_{\phi}\big( \rhoh_{\phi,0}\,\hat{\phi}(\mf x)\, \hat{\phi}(\mf x') \big)$}. In the interaction picture, the particle detectors interact with the field $\hat{\phi}(\mf x)$ according to the Hamiltonian weight\footnote{The Hamiltonian weight $\hat{h}_I(\mf x)$ (scalar) and the Hamiltonian density $\hat{\mathfrak h}_I(\mf x)$ (scalar density) are related by 
\begin{equation}
    \hat{\mathfrak h}_I(\mf x)=\sqrt{-g}\; \hat{h}_I(\mf x) \;,
\end{equation}
where $g$ is the determinant of the metric.}~\cite{us,us2}
\begin{equation} \label{eq:HamiltonianWeightAB}
    \hat{h}_I(\mf x) = \sum_{{i} \in \{ \ts a, \ts b \}} \lambda_{i} \Lambda_{i}(\mf x) \hat{\mu}_{{i}}(\tau_{{i}}(\mf x)) \hat{\phi}(\mf x),
\end{equation}
 where $\lambda_{i}$ and $\Lambda_{i}(\mf x)$ are the couping strength and the spacetime smearing of detector ${i}$, respectively. We will assume that the coupling strengths $\lambda_\ts{a}$ and $\lambda_\ts{b}$ are of the same order of magnitude, and denote their products of order $k$ as $\mathcal{O}(\lambda^k)$. Eq.~\eqref{eq:HamiltonianWeightAB} prescribes a linear coupling between the field operator and the monopole moment of each particle detector, which is given by
\begin{equation}
    \muh_{{i}}(\tau_{{i}}) = \hat{\sigma}_{{i}}^+e^{\ii\Omega_{{i}} \tau_{{i}}} + \hat{\sigma}_{{i}}^{-}e^{-\ii\Omega_{{i}} \tau_{{i}}},
\end{equation}
with $\hat{\sigma_{{i}}}^{\pm}$ being the raising and lowering ladder operators of the UDW detector. The field operator can be written as (see, e.g.,~\cite{Fulling1989,Wald2})
\begin{equation}
    \hat{{\phi}}(\mf x) = \int \dd^d \bm k \left(u_{\bm k} ( \mf x) \hat{a}_{\bm k} + u_{\bm k}^* ( \mf x) \hat{a}^\dagger_{\bm k}\right) \,,
\end{equation}
where $\{u_{\bm{k}}(\mf x),u_{\bm{k}}^*(\mf x)\}$ is a complete set of solutions of the field equation of motion, orthornormal with respect to the Klein-Gordon inner product. 

With respect to a certain time parameter $t$, the initial state $\rhoh_0$ evolves according to the following time evolution operator:
\begin{equation} \label{eq:TEO}
    \hat{U}_t = \mathcal{T}_{t}\,\text{exp}\left(-\ii \int \dd V \hat{h}_{I}(\mf x)\right),
\end{equation}
where $\mathcal{T}_t$ denotes time-ordering with respect to $t$, and \mbox{$\dd V=\sqrt{-g} \,\dd^d \mf{x}$} is the spacetime invariant volume element. Notice that if the model were covariant, it should not matter what parameter $t$ we choose for the time ordering. Indeed, if the detectors were pointlike, since they move in timelike trajectories, the time ordering of the points of interaction is unambiguous. However, if the detectors are smeared, the results can in general be different for different time parameters, something a physical model cannot allow (see~\cite{us2} for more details).


With the prescribed evolution operator, the state $\rhoh$ of the detectors-field system in the absolute causal future of the interaction region is given by
\begin{equation}
    \rhoh = \hat{U}_t \rhoh_{0} \hat{U}_t^{\dagger} \;.
\end{equation}
Since this expression does not admit a closed form in general, it is commonplace to apply perturbative techniques assuming a weak coupling $\lambda \ll 1$. We can thus Dyson expand the time evolution operator as
\begin{equation} 
    \hat{U}_t = \mathds{1} + \sum_{m \,\geq\, 1}\hat{U}_t^{(m)},
\end{equation}
where each $\hat{U}^{(m)}$ is proportional to the $m$-th power $\lambda^m$ of the coupling strength. Namely,
\begin{align} \label{eq:Dyson}
    U_t^{(m)} &= (-\ii)^m \int \dd V_1 \ldots \int  \,\dd V_m\, \,\hat{h}_I(\mf x_1) \ldots \,\hat{h}_I(\mf x_m)\, \\ \nonumber &\phantom{==}\times \theta[t(\mf x_1) -t(\mf x_2)] \ldots \theta[t(\mf x_{m-1})-t(\mf x_{m})] \label{eq:Um}.
\end{align}
 The time-evolved state can then also be expanded as
\begin{equation}
    \rhoh = \sum_{m\geq 0} \rhoh^{(m)},
\end{equation}
where
\begin{equation}\label{eq:expansionrho}
    \rhoh^{(m)} = \sum_{k = 0}^{m} \hat{U}_t^{(k)} \rhoh^{(0)} \left(\hat{U}_t^{(m-k)}\right)^{\dagger},
\end{equation}
 and in particular $\rhoh^{(0)}=\rhoh_0$. Once the interactions are switched off, we are only interested in the information extracted by the detectors. Their final density operator is obtained tracing out the field degrees of freedom
\begin{equation}\label{eq:expansionrhoab}
    \rhoh_{\textsc{ab}} = \text{Tr}_{\phi}(\,\hat{U}_t \rhoh^{(0)} \hat{U}_t^{\dagger} \,)= \rhoh_\textsc{ab}^{(0)}+\rhoh_\textsc{ab}^{(2)} + \mathcal{O}(\lambda^4).
\end{equation}
Note that since the odd-point functions of the initial field state vanish, so do the odd terms in the expansion~\eqref{eq:expansionrhoab}, and that $\rhoh^{(0)}_{\ts{ab}}=\rhoh_{\ts{ab},0}$. $\rhoh^{(2)}_{\ts{ab}}$ is given by
\begin{align}\label{eq:rhoab-2} 
    \rhoh^{(2)}_{\ts{ab}} &= \sum_{{i},{j} \in \{\ts{a},\ts{b} \}}\lambda_{i} \lambda_{j} \int \dd V \dd V' \,\Lambda_{i}(\mf x) \Lambda_{j}(\mf x') \\ 
    &\phantom{==} \times \Big[
    \muh_{i}(\tau_{i}) \rhoh_{\ts{ab}}^{(0)} \muh_{j}(\tau_{j}')\, W(\mf x', \mf x)  \nonumber \\  
    &\phantom{=====}- \muh_{i}(\tau_{i}) \muh_{j}(\tau_{j}')   \rhoh_{\ts{ab}}^{(0)} \,W(\mf x, \mf x') \theta(t-t')  \nonumber \\  
     &\phantom{=====}-  \rhoh_{\ts{ab}}^{(0)}  \muh_{j}(\tau_{j}') \muh_{i}(\tau_{i}) \,W(\mf x', \mf x) \theta(t-t') \Big]. \nonumber 
\end{align}
In this work we will be concerned with how the dependence of the evolution operator prescribed in Eq.~\eqref{eq:TEO} on the time parameter $t$ affects the prediction of the entanglement harvested by the detectors. We recall that for the model to make physical sense, its predictions cannot depend on the choice of time parameter $t$.


\subsection*{Covariance of the evolution operator} \label{covarianceIssue}

In this Subsection we discuss the covariance of the time evolution in the model for entanglement harvesting presented above. The spacetime integral in Eq. \eqref{eq:TEO} for the time-evolution operator is independent of any choice of coordinates, and therefore covariant. The time-ordering operation $\mathcal{T}_t$ is the only place where we could be introducing a dependence on the time-parameter $t$. 
This dependence does not break covariance whenever the Hamiltonian weight of the interaction is microcausal~\cite{us2} (i.e., whenever $[\,\hat{h}_I(\mf x),\hat{h}_I(\mf x')\,]=0$ for $\mf x$ and $\mf x'$ spacelike separated). This condition, fulfilled for pointlike detectors, fails to be satisfied when smeared detectors are considered.



In particular, for two time parameters $t$ and $s$, the amount of violation of covariance manifested in $\rhoh_\ts{ab}$ can be quantified by the following ``non-covariant'' contribution~\cite{us2}
\begin{align}\label{eq:nc}
    \rhoh_{\ts{ab}}^{\ts{nc}}&=\rhoh_{\ts{ab}}^{t}-\rhoh_{\ts{ab}}^{s}= \lambda^2\,a \big[\rhoh_{\ts{a},0}, \hat{\sigma}_{z,\ts a} \big] \otimes \rhoh_{\ts b,0}  \nonumber \\
    &\phantom{==\;}+ \lambda^2\,b\,\rhoh_{\ts a}^{(0)} \otimes \big[\rhoh_{\ts{b},0}, \hat{\sigma}_{z,\ts b} \big] + \mathcal{O}(\lambda^4), 
\end{align}
where $\rhoh_{\ts{ab}}^{r}=\text{Tr}_\phi\big(\hat{U}_{r} \rhoh^{(0)} \hat{U}_{r}^{\dagger}\big)$,  \mbox{$\hat{\sigma}_{z,{i}} = \proj{g_{{i}}}{g_{{i}}} - \proj{e_{{i}}}{e_{{i}}}$} for ${i} \in \{\ts{A,B}\}$, and the factors of proportionality $a$ and $b$ depend on $t$, $s$, the initial field state, and are suppressed by the smearing lengthscales of detectors $\ts{A}$ and $\ts{B}$, respectively~\cite{us2}. 

Indeed we see that the second order non-covariant terms vanish if the detectors are initially in energy eigenstates of their free Hamiltonians, but not for general initial states.




\section{Covariance of entanglement harvesting}

\subsection{General initial pure states}

First let us consider that the detectors are initially in arbitrary pure states, that is
\begin{equation}
    \rhoh_{i,0} = \proj{\psi_i}{\psi_i} \;,
\end{equation}
for $i \in \{\ts{a},\ts{b}\}$.  In this Subsection we will show that the entanglement harvested is covariant at leading order even in these cases, despite the state $\rhoh_{\ts{ab}}$ possibly having non-covariant second order terms. 


 In order to quantify the entanglement in $\rhoh_\ts{ab}$, we use the negativity~\cite{Vidal2002} (which is a faithful entanglement measure for two two-qubit systems), defined as
 \begin{equation} \label{eq:defNegativity}
    \mathcal{N}(\rhoh_{\ts{ab}}) = \sum_{j} \frac{|x_j|-x_j}{2},
\end{equation}
where $x_j$ are the eigenvalues of the partially transposed density matrix.
Since the detectors are initially in arbitrary pure states, we can pick states $\ket{\chi_{\ts a}}$ and $\ket{\chi_{\ts{b}}}$ that are orthogonal to $\ket{\psi_{\ts a}}$ and $\ket{\psi_{\ts{b}}}$, respectively. Then, we work in the basis $\{\ket{\psi_\textsc{a} \psi_{\textsc{b}}}, \ket{\psi_\textsc{a} \chi_{\textsc{b}}}, \ket{\chi_\textsc{a} \psi_{\textsc{b}}}, \ket{\chi_\textsc{a} \chi_{\textsc{b}}}\}$. Without loss of generality, we can assume that the change of basis between $\{ \ket{\psi_{{i}}}, \ket{\chi_{{i}}}\}$ and the energy eigenbasis of the $i$-th detector takes the form
\begin{align}
    \ket{g_{{i}}} &= \cos{\alpha_{{i}}} \ket{\psi_{{i}}} + e^{\ii \beta_{{i}}} \sin{\alpha_{{i}}} \ket{\chi_{{i}}}, \\
    \ket{e_{{i}}} &= -e^{-\ii \beta_{{i}}}\sin{\alpha_{{i}}} \ket{\psi_{{i}}} +  \cos{\alpha_{{i}}} \ket{\chi_{{i}}},
\end{align}
for some $\alpha_{i} \in [0,\pi)$, $\beta_{i}\in[0,2\pi)$, ${i} \in \{\ts{A},\ts{B}\}$. Using Eq.~\eqref{eq:rhoab-2}, we show in Appendix \ref{AppendixA} that the matrix form of the time-evolved state of the particle detectors is given by
\begin{align}
    \hat{\rho}_{\textsc{ab}} &= \left(
\begin{array}{cccc}
1-\mathcal{L}_\textsc{aa}^{\text{gen}}-\mathcal{L}_\textsc{bb}^{\text{gen}} & \mathcal{X}^* & \mathcal{Y}^* & (\mathcal{M}^{\text{gen}})^\ast \\
\mathcal{X} & \mathcal{L}_\textsc{bb}^{\text{gen}} & (\mathcal{L}_\textsc{ab}^{\text{gen}})^*  & 0 \\
\mathcal{Y} & \mathcal{L}_\textsc{ab}^{\text{gen}} & \mathcal{L}_\textsc{aa}^{\text{gen}} &0\\
\mathcal{M}^{\text{gen}} & 0 & 0 & 0 \\
\end{array}\right)\nonumber \\[0.2cm]
&\phantom{==} +O(\lambda^4) \;,\label{eq:matrix}
\end{align}
with
\begin{align} 
    &\mathcal{L}_{{ij}}^{\text{gen}}= \lambda_{i}\lambda_{j} \big[ \cos^2{\alpha_{{i}}}\cos^2{\alpha_{{j}}} \,\mathcal{L}_{{ij}} \nonumber\\
    &\phantom{========} - \cos^2{\alpha_{{i}}}\sin^2{\alpha_{{j}}} \,e^{-2\ii \beta_{{j}}} \nonumber \,\mathcal{P}_{{ij}}\\ &\phantom{========} - \sin^2{\alpha_{{i}}}\cos^2{\alpha_{{j}}}\,e^{2\ii \beta_{{i}}} \,\mathcal{K}_{{ij}} \nonumber\\
    &\phantom{========}+ \sin^2{\alpha_{{i}}}\sin^2{\alpha_{{j}}}\,e^{2\ii (\beta_{{i}} - \beta_{{j}} )} \,\mathcal{Q}_{{ij}} \label{eq:LijGen} \big],\\
    &\mathcal{M}^{\text{gen}}= \lambda_\ts{a}\lambda_\ts{b}\big[\cos^2{\alpha_{\ts a}}\cos^2{\alpha_{\ts b}} \,\mathcal{M}\nonumber\\
    &\phantom{=========}- \cos^2{\alpha_{\ts a}}\sin^2{\alpha_{\ts b}}\, e^{2\ii \beta_{\ts b}} \, \mathcal{R} \nonumber \\ 
    &\phantom{=========}- \sin^2{\alpha_{\ts a}}\cos^2{\alpha_{\ts b}}\,e^{2\ii \beta_{\ts a}} \,\mathcal{S} \nonumber\\
    &\phantom{=========}+ \sin^2{\alpha_{\ts a}}\sin^2{\alpha_{\ts b}}\,e^{2\ii (\beta_{\ts a} + \beta_{\ts b} )} \,\mathcal{V} \big] \label{eq:MGen},
\end{align}
 and
 \begin{align}
    \mathcal{L}_{{ij}} &= \int \dd V \dd V' \Lambda_{{i}}(\mf x) \Lambda_{{j}}(\mf x')\, W(\mf x', \mf x)\,e^{\ii\left( \Omega_{{i}} \tau_{{i}} - \Omega_{{j}} \tau'_{{j}}\right)}, \label{eqLij} \\
    \mathcal{P}_{{ij}} &= \int \dd V \dd V' \Lambda_{{i}}(\mf x) \Lambda_{{j}}(\mf x')\, W(\mf x', \mf x)\,e^{\ii\left( \Omega_{{i}} \tau_{{i}} + \Omega_{{j}} \tau'_{{j}}\right)}, \\
    \mathcal{K}_{{ij}} &= \int \dd V \dd V' \Lambda_{{i}}(\mf x) \Lambda_{{j}}(\mf x')\, W(\mf x', \mf x)\,e^{-\ii\left( \Omega_{{i}} \tau_{{i}} + \Omega_{{j}} \tau'_{{j}}\right)},  \\
    \mathcal{Q}_{{ij}} &= \int \dd V \dd V' \Lambda_{{i}}(\mf x) \Lambda_{{j}}(\mf x')\, W(\mf x', \mf x)\,e^{\ii\left( -\Omega_{{i}} \tau_{{i}} + \Omega_{{j}} \tau'_{{j}}\right)}, \\
    \mathcal{M} &= - \int \dd V \dd V' \Lambda_{\ts a}(\mf x) \Lambda_{\ts b}(\mf x')\, G_{F}(\mf x, \mf x')\,e^{\ii\left( \Omega_{\ts a} \tau_{\ts a} + \Omega_{\ts b} \tau'_{\ts b}\right)}, \label{eqM} \\
    \mathcal{R} &= - \int \dd V \dd V' \Lambda_{\ts a}(\mf x) \Lambda_{\ts b}(\mf x')\, G_{F}(\mf x, \mf x')\,e^{\ii\left( \Omega_{\ts a} \tau_{\ts a} - \Omega_{\ts b} \tau'_{\ts b}\right)},  \label{eqR}\\
    \mathcal{S} &= - \int \dd V \dd V' \Lambda_{\ts a}(\mf x) \Lambda_{\ts b}(\mf x')\, G_{F}(\mf x, \mf x')\,e^{\ii\left( -\Omega_{\ts a} \tau_{\ts a} + \Omega_{\ts b} \tau'_{\ts b}\right)},  \label{eqS}\\
    \mathcal{V} &= - \int \dd V \dd V' \Lambda_{\ts a}(\mf x) \Lambda_{\ts b}(\mf x')\, G_{F}(\mf x, \mf x')\,e^{-\ii\left( \Omega_{\ts a} \tau_{\ts a} + \Omega_{\ts b} \tau'_{\ts b}\right)}. \label{eqV}
 \end{align}
The expressions for $\mathcal{X}$ and $\mathcal{Y}$ are more cumbersome and we leave them for Appendix \ref{AppendixA}. In Eqs.~\eqref{eqM}--\eqref{eqV}, $G_{F}(\mf x, \mf x')$ is the Feynman propagator, given by
 \begin{equation}
     G_{F}(\mf x, \mf x') = \theta(t-t')W(\mf x, \mf x')   +\theta(t'-t)W(\mf x', \mf x).
 \end{equation}
 Notice that the Feynman propagator is independent of the time parameter $t$ used for the time-ordering operator $\mathcal{T}_t$. This is because:
\begin{enumerate}
    \item If $\mf x$ and $\mf x'$ are causally connected, any notion of time-ordering respects their causal order. If, for instance, $\mf x'$ precedes $\mf x$, then for any time parameter we will have that $t > t'$, and thus $\theta(t-t') = 1$ and $\theta(t'-t) = 0$.
    \item If $\mf x$ and $\mf x'$ are spacelike separated, the Wightman function satisfies $W(\mf x, \mf x') = W(\mf x', \mf x)$. This implies 
    \begin{equation}
        G_{F}(\mf x, \mf x') = W(\mf x, \mf x')
    \end{equation}
\end{enumerate}
Therefore, the terms $\mathcal{L}^{gen}_{{ij}}$ and $\mathcal{M}^{gen}$ are covariant, since they do not depend on the time-coordinate $t$. In contrast, as shown in Appendix~\ref{AppendixA}, the terms $\mathcal{X}$ and $\mathcal{Y}$ are not 
covariant in general, since they depend on the ordering of $t$ and $t'$.

Now, given the form for $\rhoh_\ts{ab}$ given in Eq.~\eqref{eq:matrix}, we can conclude that the negativity is given by
\begin{equation}\label{negativity}
    \mathcal{N}(\rhoh_\ts{ab})=\max (0,\mathcal{N}^{(2)})+\mathcal{O}(\lambda^4) \;,
\end{equation}
with
\begin{equation}\label{negativitytilde}
    \mathcal{N}^{(2)}=\sqrt{\frac{(\L_{\ts{aa}}^{\text{gen}} - \L_{\ts{bb}}^{\text{gen}})^2}{4}+ |\mathcal{M}^{\text{gen}}|^2}\, -\,\frac{\L_{\ts{aa}}^{\text{gen}} +\L_{\ts{bb}}^{\text{gen}}}{2}.
\end{equation}
This proves that the negativity is covariant at leading order. And that is despite the joint state of the detectors displaying in general non-covariant terms at second order in perturbation theory, which is the leading order for the entanglement harvested.

It is worth mentioning that Eqs.~\eqref{negativity} and~\eqref{negativitytilde}, along with Eqs.~\eqref{eq:LijGen}--\eqref{eqV}, generalize the expression found in~\cite{Pozas-Kerstjens:2015} for the case in which the detectors are initially in their ground states. Notice that that the negativity takes the same functional form in this case, with the replacements \mbox{$\mathcal{L}_{\ts{aa}} \rightarrow \mathcal{L}_{\ts{aa}}^{\text{gen}}$}, \mbox{$\mathcal{L}_{\ts{bb}} \rightarrow \mathcal{L}_{\ts{bb}}^{\text{gen}}$}, and \mbox{$\mathcal{M} \rightarrow \mathcal{M}^{\text{gen}}$}.

Our results also generalize and reinforce the interpretation of entanglement harvesting as a competition between local noise and non-local correlations, which has been pointed out for detectors initially in their ground states since the first works on entanglement harvesting. Indeed, Eq.~\eqref{negativitytilde} shows that negativity is a competition between the non-local term $\mathcal{M}$ and the local terms $\mathcal{L}_{\ts{aa}}^{\text{gen}}$ and $\mathcal{L}_{\ts{bb}}^{\text{gen}}$: it is straightforward to verify that $\mathcal{N}^{(2)} \leq |\mathcal{M}^{\text{gen}}|$, and therefore 
$\mathcal{L}_{\ts{aa}}^{\text{gen}}$ and $\mathcal{L}_{\ts{bb}}^{\text{gen}}$ act as local noise that can only decrease the entanglement harvested. 

\vspace{5mm}
\subsection{Low mixedness states}

So far, we have only analyzed the case in which the detectors are initially in pure states. Mixedness present in the initial states of the detectors has been shown to rapidly diminish the ability of the detectors to harvest entanglement to almost zero~\cite{BrunoDan}. Therefore the most important states for which we would like to analyze the covariance of  the models of entanglement harvesting are those who only have a small degree of mixedness.

Let us consider that the detectors are initially in mixed states with a small degree of mixedness in comparison with the coupling parameter. Namely, we consider here initial states
\begin{align} \label{eq:rho0mixed}
    \rhoh_{\ts a,0} &= (1-p_{\ts a}) \ketbra{\psi_{\ts a}}{\psi_{\ts a}} + p_{\ts a} \ketbra{\chi_{\ts a}}{\chi_{\ts a}}, \\
    \rhoh_{\ts b,0} &= (1-p_{\ts b}) \ketbra{\psi_{\ts b}}{\psi_{\ts b}} + p_{\ts b} \ketbra{\chi_{\ts b}}{\chi_{\ts b}},
\end{align}
with $1\gg p_{\ts a},p_{\ts b}\sim \mathcal{L}_{ij}^{\text{gen}}$. In this case, the joint state of the detectors after the entanglement harvesting protocol has taken place can be readily computed with the same techniques used in Appendix~\ref{AppendixA}. We take into account that the terms involving $\ket{\chi_i}$ are weighted by $p_i\sim \mathcal{L}_{ij}^{\text{gen}}$ in the initial state, and therefore are already second order in perturbation theory. Using the orthogonality $\braket{\psi_i}{\chi_i}=0$ we see that all corrections appear on the diagonal terms. The final density matrix in this case is given by 
\begin{widetext}
\begin{equation}
    \hat{\rho}_{\textsc{ab}}^{\text{gen}} = \left(
\begin{array}{cccc}
1-\mathcal{L}_\textsc{aa}^{\text{gen}}-\mathcal{L}_\textsc{bb}^{\text{gen}} - p_{\ts a} - p_{\ts b} &  \mathcal{X}^*& \mathcal{Y}^* & (\mathcal{M}^{\text{gen}})^\ast \\
 \mathcal{X} & \mathcal{L}_\textsc{bb}^{\text{gen}} + p_{\ts b} & (\mathcal{L}_\textsc{ab}^{\text{gen}})^\ast & 0 \\
\mathcal{Y}& \mathcal{L}_\textsc{ab}^{\text{gen}} & \mathcal{L}_\textsc{aa}^{\text{gen}} +  p_{\ts a} & 0 \\
\mathcal{M}^{\text{gen}} & 0 & 0 & 0 \\
\end{array}\right)+O(\lambda^4)\;.\\
\end{equation}
The mixedness in each particle detector acts as a source of additional noise that adds up to the local transition probabilities $\mathcal{L}_{\ts{aa}}^{\text{gen}}$ and $\mathcal{L}_{\ts{bb}}^{\text{gen}}$. The negativity in this case is given by
\begin{equation} \label{eq:negativityMixed}
    \mathcal{N} = \text{max}\left(\sqrt{\frac{(\L_{\ts{aa}}^{\text{gen}} + p_{\ts a} - \L_{\ts{bb}}^{\text{gen}} - p_{\ts b})^2}{4}+ |\mathcal{M}^{\text{gen}}|^2}\, -\,\frac{\L_{\ts{aa}}^{\text{gen}} +p_{\ts a} +\L_{\ts{bb}}^{\text{gen}} + p_{\ts b}}{2},0\right) +\mathcal{O}(\lambda^4),
\end{equation}
which is simply the expression given in Eqs.~\eqref{negativity} and~\eqref{negativitytilde} with the substitutions \mbox{$\mathcal{L}_{\ts{aa}}^{\text{gen}} \rightarrow \mathcal{L}_{\ts{aa}}^{\text{gen}}+p_\ts{a}$}, and \mbox{$\mathcal{L}_{\ts{bb}}^{\text{gen}} \rightarrow \mathcal{L}_{\ts{bb}}^{\text{gen}}+p_\ts{b}$}.
\end{widetext}
Eq.~\eqref{eq:negativityMixed} is covariant, since the mixedness only modifies the local noise terms in a way independent of the time-ordering parameter $t$. We also see that (as discussed in \cite{BrunoDan}) mixedness competes against entanglement harvesting by increasing the local noise.  


\section{Conclusions}

We have generalized the protocol of entanglement harvesting to arbitrary initial pure states of the detectors as well as non-pure initial states with low mixedness for any initial quasifree state of the field. These results can be applied to specific experimental setups to analyze the effect of the initial states on the entanglement harvested, which might play an important role in the optimization of entanglement extraction.

We have also shown that despite of the fact that smeared particle detector models suffer from violations of covariance at second order in perturbation theory for arbitrary initial states, those violations of covariance do not impact the modelling of entanglement harvesting protocols. The result guarantees that we can safely use particle detectors to probe the entanglement structure of quasifree field states at leading order, regardless of the initial setup of the detectors.

\section{Acknowledgements}

HMG has been partially
supported by the mobility grants program of Centre de Formació
Interdisciplinària Superior (CFIS) - Universitat Politècnica de
Catalunya (UPC). JPG is supported by a Mike and Ophelia Lazaridis Fellowship. JPG also received the support of a fellowship from ``La Caixa'' Foundation (ID 100010434, with fellowship code LCF/BQ/AA20/11820043). EMM acknowledges support through the Discovery Grant Program of the Natural Sciences and Engineering Research Council of Canada (NSERC). EMM also acknowledges support of his Ontario Early Researcher award. Research at Perimeter Institute is supported in part by the Government of Canada through the Department of Innovation, Science and Industry Canada and by the Province of Ontario through the Ministry of Colleges and Universities.

\onecolumngrid
\appendix
\section{Time-evolved state} \label{AppendixA}

In this Appendix we show that in the basis $\{\ket{\psi_\textsc{a} \psi_{\textsc{b}}}, \ket{\psi_\textsc{a} \chi_{\textsc{b}}}, \ket{\chi_\textsc{a} \psi_{\textsc{b}}}, \ket{\chi_\textsc{a} \chi_{\textsc{b}}}\}$, the matrix form of the time-evolved state $\rhoh_{\ts{ab}}$ is given by Eq. \eqref{eq:matrix}. The matrix form of the initial state of the detectors in this basis is simply
\begin{equation}
    \rhoh_{\ts{ab},0} = \proj{\psi_{\ts{a}}}{\psi_{\ts a}} \otimes \proj{\psi_{\ts{b}}}{\psi_{\ts b}} =\left(
\begin{array}{cccc}
1 & 0 & 0 & 0\\
0 &0&0 & 0 \\
0 & 0&0 & 0 \\
0 & 0 & 0 & 0 \\
\end{array}\right).\\
 \end{equation}
 The second order terms in the coupling parameter are given by Eq. \eqref{eq:rhoab-2}. In this equation, we can separate the local and the non local terms as follows:
 \begin{equation} \label{eq:A2}
     \rhoh_{\ts{ab}}^{(2)} = \hatr^{(2)}_{\ts a} \otimes \rhoh_{\ts{b},0} + \rhoh_{\ts{a},0} \otimes \hatr^{(2)}_{\ts a} +  \hatr_{\ts{ab}}^{(2)}
 \end{equation}
 The local terms are given by
\begin{align} 
    \hatr^{(2)}_{{i}}&= \lambda_{i}^2 \int \dd V \dd V' \,\Lambda_{i}(\mf x) \Lambda_{j}(\mf x')\bigg[
    \muh_{i}(\tau_{i}) \rhoh_{{i},0} \muh_{i}(\tau_{i}')\, W(\mf x', \mf x) - \muh_{i}(\tau_{i}) \muh_{i}(\tau_{i}')   \rhoh_{{i},0} \,W(\mf x, \mf x') \theta(t-t') \\
    &\phantom{=============================}-  \rhoh_{{i},0}  \muh_{i}(\tau_{i}') \muh_{i}(\tau_{i}) \,W(\mf x', \mf x) \theta(t-t') \bigg], \nonumber
\end{align}
for ${i} \in \{ \ts{A}, \ts{B}\}$. In the basis $\{\ket{\psi_{{i}}}, \ket{\chi_{{i}}}\}$, the local terms take the form
 \begin{equation}
     \hatr_{{i}}^{(2)} = \left(
\begin{array}{cc}
-\mathcal{L}_{\ts{ii}}^{\text{gen}} & \mathcal{I}_{\ts{ii}}^*  \\
\mathcal{I}_{\ts{ii}} & \mathcal{L}_{\ts{ii}}^{\text{gen}} \\
\end{array}\right),
 \end{equation}
 with $\mathcal{L}_{\ts{ii}}^{\text{gen}}$ given by Eq. \eqref{eq:LijGen}, and
\begin{align}
     \mathcal{I}_{\ts{ii}} &= 
     \cos^3{\alpha_{{i}}} \sin{\alpha_{{i}}} \left(-e^{-\ii \beta_{{i}}}\mathcal{S}_{\ts{ii}} +e^{\ii \beta_{{i}}}(-\mathcal{L}_{\ts{ii}}+\eta_{{i}}-\gamma_{{i}}) \right)+ 
     \cos{\alpha_{{i}}} \sin^3{\alpha_{{i}}} \left(e^{3\ii \beta_{{i}}}\mathcal{K}_{\ts{ii}} +e^{\ii \beta_{{i}}}(\mathcal{R}_{\ts{ii}}+\eta_{i}-\gamma_{i}) \right),
\end{align}
with 
\begin{align}
     \gamma_{{i}} &= \lambda_{{i}}^2 \int \dd V \dd V' \Lambda_{{i}}(\mf x) \Lambda_{{i}}(\mf x')\, W(\mf x', \mf x)\,e^{\ii\Omega_{{i}}\left(  \tau_{{i}} -  \tau_{{i}}'\right)}\theta(t-t'), \\
    \eta_{{i}} &= \lambda_{{i}}^2 \int \dd V \dd V' \Lambda_{{i}}(\mf x) \Lambda_{{i}}(\mf x')\, W(\mf x', \mf x)\,e^{-\ii\Omega_{{i}}\left(  \tau_{{i}} -  \tau_{{i}}'\right)}\theta(t-t'). 
 \end{align}
 Notice that $\gamma_{i}$ and $\eta_{i}$ depend on the time-ordering parameter $t$, and therefore they are not covariant in general. 

The non-local terms $\hatr_{\ts{ab}}$ are given by the same expression as in Eq. \eqref{eq:rhoab-2}, but considering only the terms with ${i} \neq {j}$. In the basis $\{\ket{\psi_\textsc{a} \psi_{\textsc{b}}}, \ket{\psi_\textsc{a} \chi_{\textsc{b}}}, \ket{\chi_\textsc{a} \psi_{\textsc{b}}}, \ket{\chi_\textsc{a} \chi_{\textsc{b}}}\}$, they yield 
 \begin{align}
    \hatr_{\textsc{ab}}^{(2)} &= \left(
\begin{array}{cccc}
0 & \mathcal{J}_1^* & \mathcal{J}_2^* & (\mathcal{M}^{\text{gen}})^\ast \\
\mathcal{J}_1 & 0 & (\mathcal{L}_\textsc{ab}^{\text{gen}})^*  & 0 \\
\mathcal{J}_2 & \mathcal{L}_\textsc{ab}^{\text{gen}} & 0 &0\\
\mathcal{M}^{\text{gen}} & 0 & 0 & 0 \\
\end{array}\right) \;,
\end{align}
 where $\mathcal{M}^{\text{gen}}$ and $\mathcal{L}_{\ts{ab}}^{\text{gen}}$ are given in Eqs. \eqref{eq:LijGen} and \eqref{eq:MGen}, and $\mathcal{J}_1$ and $\mathcal{J}_2$ read
\begin{align}
     \mathcal{J}_1 &= \frac{1}{2}e^{-\ii \beta_{\ts a}}\sin{2 \alpha_{\ts a}}\bigg[\big(-\mathcal{M} + e^{2 \ii \beta_{\ts a}} (-\mathcal{S} - \mathcal{L}_{\ts{ab}}^*) - \mathcal{K}_{\ts{ab}}^*
     \big) \cos^2{\alpha_{\ts b}} + 
  e^{2 \ii \beta_{\ts b}}
     \big(\mathcal{R} + e^{2\ii \beta_{\ts a}} (\mathcal{V} + \mathcal{P}_{\ts{ab}}^*) + 
     \mathcal{Q}_{\ts{ab}}^*\big) \sin^2{\alpha_{\ts b}} \bigg] \;, \\
     \mathcal{J}_2 &= \frac{1}{2}e^{-\ii \beta_{\ts b}}\sin{2 \alpha_{\ts b}}\bigg[\big(-\mathcal{M} + e^{2 \ii \beta_{\ts b}} (-\mathcal{R} - \mathcal{L}_{\ts{ab}}) - \mathcal{P}_{\ts{ab}}^*
     \big) \cos^2{\alpha_{\ts b}} +  e^{2 \ii \beta_{\ts a}}
     \big(\mathcal{S} + e^{2\ii \beta_{\ts a}} (\mathcal{V} + \mathcal{K}_{\ts{ab}}) + 
     \mathcal{Q}_{\ts{ab}}\big) \sin^2{\alpha_{\ts a}} \bigg] \;.
 \end{align}
Adding the contributions of the local and the non-local terms in Eq. \eqref{eq:A2} and using $\rhoh_{\ts{ab}} = \rhoh_{\ts{ab}}^{(0)}+\rhoh_{\ts{ab}}^{(2)}+\mathcal{O}(\lambda^4)$, we recover the matrix form given in Eq. \eqref{eq:matrix}, with the definitions
 \begin{align}
     \mathcal{X} = \mathcal{I}_{\ts{bb}} + \mathcal{J}_1, \\
      \mathcal{Y} = \mathcal{I}_{\ts{aa}} + \mathcal{J}_2.
 \end{align}
 Observe that the non-covariance of the terms $\mathcal{X}$ and $\mathcal{Y}$ is due to the local terms $\mathcal{I}_{\ts{aa}}$ and $\mathcal{I}_{\ts{bb}}$, since $\mathcal{J}_1$ and $\mathcal{J}_2$ are independent of the time parameter $t$.

\twocolumngrid
\bibliography{references}

\begin{thebibliography}{42}%
\makeatletter
\providecommand \@ifxundefined [1]{%
 \@ifx{#1\undefined}
}%
\providecommand \@ifnum [1]{%
 \ifnum #1\expandafter \@firstoftwo
 \else \expandafter \@secondoftwo
 \fi
}%
\providecommand \@ifx [1]{%
 \ifx #1\expandafter \@firstoftwo
 \else \expandafter \@secondoftwo
 \fi
}%
\providecommand \natexlab [1]{#1}%
\providecommand \enquote  [1]{``#1''}%
\providecommand \bibnamefont  [1]{#1}%
\providecommand \bibfnamefont [1]{#1}%
\providecommand \citenamefont [1]{#1}%
\providecommand \href@noop [0]{\@secondoftwo}%
\providecommand \href [0]{\begingroup \@sanitize@url \@href}%
\providecommand \@href[1]{\@@startlink{#1}\@@href}%
\providecommand \@@href[1]{\endgroup#1\@@endlink}%
\providecommand \@sanitize@url [0]{\catcode `\\12\catcode `\$12\catcode
  `\&12\catcode `\#12\catcode `\^12\catcode `\_12\catcode `\%12\relax}%
\providecommand \@@startlink[1]{}%
\providecommand \@@endlink[0]{}%
\providecommand \url  [0]{\begingroup\@sanitize@url \@url }%
\providecommand \@url [1]{\endgroup\@href {#1}{\urlprefix }}%
\providecommand \urlprefix  [0]{URL }%
\providecommand \Eprint [0]{\href }%
\providecommand \doibase [0]{https://doi.org/}%
\providecommand \selectlanguage [0]{\@gobble}%
\providecommand \bibinfo  [0]{\@secondoftwo}%
\providecommand \bibfield  [0]{\@secondoftwo}%
\providecommand \translation [1]{[#1]}%
\providecommand \BibitemOpen [0]{}%
\providecommand \bibitemStop [0]{}%
\providecommand \bibitemNoStop [0]{.\EOS\space}%
\providecommand \EOS [0]{\spacefactor3000\relax}%
\providecommand \BibitemShut  [1]{\csname bibitem#1\endcsname}%
\let\auto@bib@innerbib\@empty
\bibitem [{\citenamefont {Summers}\ and\ \citenamefont
  {Werner}(1985)}]{SUMMERS}%
  \BibitemOpen
  \bibfield  {author} {\bibinfo {author} {\bibfnamefont {S.~J.}\ \bibnamefont
  {Summers}}\ and\ \bibinfo {author} {\bibfnamefont {R.}~\bibnamefont
  {Werner}},\ }\bibfield  {title} {\bibinfo {title} {The vacuum violates
  {B}ell's inequalities},\ }\href
  {https://doi.org/https://doi.org/10.1016/0375-9601(85)90093-3} {\bibfield
  {journal} {\bibinfo  {journal} {Phys. Lett. A}\ }\textbf {\bibinfo {volume}
  {110}},\ \bibinfo {pages} {257} (\bibinfo {year} {1985})}\BibitemShut
  {NoStop}%
\bibitem [{\citenamefont {Higuchi}\ \emph {et~al.}(2017)\citenamefont
  {Higuchi}, \citenamefont {Iso}, \citenamefont {Ueda},\ and\ \citenamefont
  {Yamamoto}}]{vacuumEntanglement}%
  \BibitemOpen
  \bibfield  {author} {\bibinfo {author} {\bibfnamefont {A.}~\bibnamefont
  {Higuchi}}, \bibinfo {author} {\bibfnamefont {S.}~\bibnamefont {Iso}},
  \bibinfo {author} {\bibfnamefont {K.}~\bibnamefont {Ueda}},\ and\ \bibinfo
  {author} {\bibfnamefont {K.}~\bibnamefont {Yamamoto}},\ }\bibfield  {title}
  {\bibinfo {title} {Entanglement of the vacuum between left, right, future,
  and past: The origin of entanglement-induced quantum radiation},\ }\href
  {https://doi.org/10.1103/PhysRevD.96.083531} {\bibfield  {journal} {\bibinfo
  {journal} {Phys. Rev. D}\ }\textbf {\bibinfo {volume} {96}},\ \bibinfo
  {pages} {083531} (\bibinfo {year} {2017})}\BibitemShut {NoStop}%
\bibitem [{\citenamefont {Bombelli}\ \emph {et~al.}(1986)\citenamefont
  {Bombelli}, \citenamefont {Koul}, \citenamefont {Lee},\ and\ \citenamefont
  {Sorkin}}]{Bombelli1986}%
  \BibitemOpen
  \bibfield  {author} {\bibinfo {author} {\bibfnamefont {L.}~\bibnamefont
  {Bombelli}}, \bibinfo {author} {\bibfnamefont {R.~K.}\ \bibnamefont {Koul}},
  \bibinfo {author} {\bibfnamefont {J.}~\bibnamefont {Lee}},\ and\ \bibinfo
  {author} {\bibfnamefont {R.~D.}\ \bibnamefont {Sorkin}},\ }\bibfield  {title}
  {\bibinfo {title} {Quantum source of entropy for black holes},\ }\href
  {https://doi.org/10.1103/PhysRevD.34.373} {\bibfield  {journal} {\bibinfo
  {journal} {Phys. Rev. D}\ }\textbf {\bibinfo {volume} {34}},\ \bibinfo
  {pages} {373} (\bibinfo {year} {1986})}\BibitemShut {NoStop}%
\bibitem [{\citenamefont {Witten}(2018)}]{witten}%
  \BibitemOpen
  \bibfield  {author} {\bibinfo {author} {\bibfnamefont {E.}~\bibnamefont
  {Witten}},\ }\bibfield  {title} {\bibinfo {title} {{APS} {M}edal for
  {E}xceptional {A}chievement in {R}esearch: {I}nvited article on entanglement
  properties of quantum field theory},\ }\href
  {https://doi.org/10.1103/RevModPhys.90.045003} {\bibfield  {journal}
  {\bibinfo  {journal} {Rev. Mod. Phys.}\ }\textbf {\bibinfo {volume} {90}},\
  \bibinfo {pages} {045003} (\bibinfo {year} {2018})}\BibitemShut {NoStop}%
\bibitem [{\citenamefont {Calabrese}\ and\ \citenamefont
  {Cardy}(2004)}]{Calabrese2004}%
  \BibitemOpen
  \bibfield  {author} {\bibinfo {author} {\bibfnamefont {P.}~\bibnamefont
  {Calabrese}}\ and\ \bibinfo {author} {\bibfnamefont {J.}~\bibnamefont
  {Cardy}},\ }\bibfield  {title} {\bibinfo {title} {Entanglement entropy and
  quantum field theory},\ }\href
  {https://doi.org/10.1088/1742-5468/2004/06/p06002} {\bibfield  {journal}
  {\bibinfo  {journal} {J. Stat. Mech. Theory Exp.}\ }\textbf {\bibinfo
  {volume} {2004}},\ \bibinfo {pages} {P06002} (\bibinfo {year}
  {2004})}\BibitemShut {NoStop}%
\bibitem [{\citenamefont {Ryu}\ and\ \citenamefont
  {Takayanagi}(2006)}]{Ryu2006}%
  \BibitemOpen
  \bibfield  {author} {\bibinfo {author} {\bibfnamefont {S.}~\bibnamefont
  {Ryu}}\ and\ \bibinfo {author} {\bibfnamefont {T.}~\bibnamefont
  {Takayanagi}},\ }\bibfield  {title} {\bibinfo {title} {Holographic
  {D}erivation of {E}ntanglement {E}ntropy from the anti--de {S}itter
  {S}pace/{C}onformal {F}ield {T}heory {C}orrespondence},\ }\href
  {https://doi.org/10.1103/PhysRevLett.96.181602} {\bibfield  {journal}
  {\bibinfo  {journal} {Phys. Rev. Lett.}\ }\textbf {\bibinfo {volume} {96}},\
  \bibinfo {pages} {181602} (\bibinfo {year} {2006})}\BibitemShut {NoStop}%
\bibitem [{\citenamefont {Calabrese}\ and\ \citenamefont
  {Cardy}(2009)}]{Calabrese2009}%
  \BibitemOpen
  \bibfield  {author} {\bibinfo {author} {\bibfnamefont {P.}~\bibnamefont
  {Calabrese}}\ and\ \bibinfo {author} {\bibfnamefont {J.}~\bibnamefont
  {Cardy}},\ }\bibfield  {title} {\bibinfo {title} {Entanglement entropy and
  conformal field theory},\ }\href
  {https://doi.org/10.1088/1751-8113/42/50/504005} {\bibfield  {journal}
  {\bibinfo  {journal} {J. Phys. A Math. Theor.}\ }\textbf {\bibinfo {volume}
  {42}},\ \bibinfo {pages} {504005} (\bibinfo {year} {2009})}\BibitemShut
  {NoStop}%
\bibitem [{\citenamefont {Saravani}\ \emph {et~al.}(2014)\citenamefont
  {Saravani}, \citenamefont {Sorkin},\ and\ \citenamefont
  {Yazdi}}]{Saravani2014}%
  \BibitemOpen
  \bibfield  {author} {\bibinfo {author} {\bibfnamefont {M.}~\bibnamefont
  {Saravani}}, \bibinfo {author} {\bibfnamefont {R.~D.}\ \bibnamefont
  {Sorkin}},\ and\ \bibinfo {author} {\bibfnamefont {Y.~K.}\ \bibnamefont
  {Yazdi}},\ }\bibfield  {title} {\bibinfo {title} {Spacetime entanglement
  entropy in 1 + 1 dimensions},\ }\href
  {https://doi.org/10.1088/0264-9381/31/21/214006} {\bibfield  {journal}
  {\bibinfo  {journal} {\, Class. Quantum Gravity}\ }\textbf {\bibinfo {volume}
  {31}},\ \bibinfo {pages} {214006} (\bibinfo {year} {2014})}\BibitemShut
  {NoStop}%
\bibitem [{\citenamefont {Unruh}(1976)}]{Unruh1976}%
  \BibitemOpen
  \bibfield  {author} {\bibinfo {author} {\bibfnamefont {W.~G.}\ \bibnamefont
  {Unruh}},\ }\bibfield  {title} {\bibinfo {title} {Notes on black-hole
  evaporation},\ }\href {https://doi.org/10.1103/PhysRevD.14.870} {\bibfield
  {journal} {\bibinfo  {journal} {Phys. Rev. D}\ }\textbf {\bibinfo {volume}
  {14}},\ \bibinfo {pages} {870} (\bibinfo {year} {1976})}\BibitemShut
  {NoStop}%
\bibitem [{\citenamefont {DeWitt}(1980)}]{DeWitt}%
  \BibitemOpen
  \bibfield  {author} {\bibinfo {author} {\bibfnamefont {B.}~\bibnamefont
  {DeWitt}},\ }\href@noop {} {\emph {\bibinfo {title} {General Relativity; an
  Einstein Centenary Survey}}}\ (\bibinfo  {publisher} {Cambridge University
  Press},\ \bibinfo {address} {Cambridge, UK},\ \bibinfo {year}
  {1980})\BibitemShut {NoStop}%
\bibitem [{\citenamefont {Valentini}(1991)}]{Valentini1991}%
  \BibitemOpen
  \bibfield  {author} {\bibinfo {author} {\bibfnamefont {A.}~\bibnamefont
  {Valentini}},\ }\bibfield  {title} {\bibinfo {title} {Non-local correlations
  in quantum electrodynamics},\ }\href
  {https://doi.org/http://dx.doi.org/10.1016/0375-9601(91)90952-5} {\bibfield
  {journal} {\bibinfo  {journal} {Phys. Lett. A}\ }\textbf {\bibinfo {volume}
  {153}},\ \bibinfo {pages} {321 } (\bibinfo {year} {1991})}\BibitemShut
  {NoStop}%
\bibitem [{\citenamefont {Reznik}(2003)}]{Reznik2003}%
  \BibitemOpen
  \bibfield  {author} {\bibinfo {author} {\bibfnamefont {B.}~\bibnamefont
  {Reznik}},\ }\bibfield  {title} {\bibinfo {title} {Entanglement from the
  {V}acuum},\ }\href {https://doi.org/10.1023/A:1022875910744} {\bibfield
  {journal} {\bibinfo  {journal} {Found. Phys.}\ }\textbf {\bibinfo {volume}
  {33}},\ \bibinfo {pages} {167} (\bibinfo {year} {2003})}\BibitemShut
  {NoStop}%
\bibitem [{\citenamefont {Reznik}\ \emph {et~al.}(2005)\citenamefont {Reznik},
  \citenamefont {Retzker},\ and\ \citenamefont {Silman}}]{Reznik1}%
  \BibitemOpen
  \bibfield  {author} {\bibinfo {author} {\bibfnamefont {B.}~\bibnamefont
  {Reznik}}, \bibinfo {author} {\bibfnamefont {A.}~\bibnamefont {Retzker}},\
  and\ \bibinfo {author} {\bibfnamefont {J.}~\bibnamefont {Silman}},\
  }\bibfield  {title} {\bibinfo {title} {Violating {B}ell's inequalities in
  vacuum},\ }\href {http://link.aps.org/abstract/PRA/v71/e042104} {\bibfield
  {journal} {\bibinfo  {journal} {Phys. Rev. A}\ }\textbf {\bibinfo {volume}
  {71}},\ \bibinfo {eid} {042104} (\bibinfo {year} {2005})}\BibitemShut
  {NoStop}%
\bibitem [{\citenamefont {Steeg}\ and\ \citenamefont {Menicucci}(2009)}]{Nick}%
  \BibitemOpen
  \bibfield  {author} {\bibinfo {author} {\bibfnamefont {G.~V.}\ \bibnamefont
  {Steeg}}\ and\ \bibinfo {author} {\bibfnamefont {N.~C.}\ \bibnamefont
  {Menicucci}},\ }\bibfield  {title} {\bibinfo {title} {Entangling power of an
  expanding universe},\ }\href {https://doi.org/10.1103/PhysRevD.79.044027}
  {\bibfield  {journal} {\bibinfo  {journal} {Phys. Rev. D}\ }\textbf {\bibinfo
  {volume} {79}},\ \bibinfo {pages} {044027} (\bibinfo {year}
  {2009})}\BibitemShut {NoStop}%
\bibitem [{\citenamefont {Pozas-Kerstjens}\ and\ \citenamefont
  {Mart\'{i}n-Mart\'{i}nez}(2015)}]{Pozas-Kerstjens:2015}%
  \BibitemOpen
  \bibfield  {author} {\bibinfo {author} {\bibfnamefont {A.}~\bibnamefont
  {Pozas-Kerstjens}}\ and\ \bibinfo {author} {\bibfnamefont {E.}~\bibnamefont
  {Mart\'{i}n-Mart\'{i}nez}},\ }\bibfield  {title} {\bibinfo {title}
  {Harvesting correlations from the quantum vacuum},\ }\href
  {https://doi.org/10.1103/PhysRevD.92.064042} {\bibfield  {journal} {\bibinfo
  {journal} {Phys. Rev. D}\ }\textbf {\bibinfo {volume} {92}},\ \bibinfo
  {pages} {064042} (\bibinfo {year} {2015})}\BibitemShut {NoStop}%
\bibitem [{\citenamefont {Silman}\ and\ \citenamefont
  {Reznik}(2007)}]{Reznik2}%
  \BibitemOpen
  \bibfield  {author} {\bibinfo {author} {\bibfnamefont {J.}~\bibnamefont
  {Silman}}\ and\ \bibinfo {author} {\bibfnamefont {B.}~\bibnamefont
  {Reznik}},\ }\bibfield  {title} {\bibinfo {title} {Long-range entanglement in
  the {D}irac vacuum},\ }\href {https://doi.org/10.1103/PhysRevA.75.052307}
  {\bibfield  {journal} {\bibinfo  {journal} {Phys. Rev. A}\ }\textbf {\bibinfo
  {volume} {75}},\ \bibinfo {pages} {052307} (\bibinfo {year}
  {2007})}\BibitemShut {NoStop}%
\bibitem [{\citenamefont {Mart{\'{\i}}n-Mart{\'{\i}}nez}\ and\ \citenamefont
  {Menicucci}(2012)}]{Edu2012}%
  \BibitemOpen
  \bibfield  {author} {\bibinfo {author} {\bibfnamefont {E.}~\bibnamefont
  {Mart{\'{\i}}n-Mart{\'{\i}}nez}}\ and\ \bibinfo {author} {\bibfnamefont
  {N.~C.}\ \bibnamefont {Menicucci}},\ }\bibfield  {title} {\bibinfo {title}
  {Cosmological quantum entanglement},\ }\href
  {https://doi.org/10.1088/0264-9381/29/22/224003} {\bibfield  {journal}
  {\bibinfo  {journal} {Class. Quantum Gravity}\ }\textbf {\bibinfo {volume}
  {29}},\ \bibinfo {pages} {224003} (\bibinfo {year} {2012})}\BibitemShut
  {NoStop}%
\bibitem [{\citenamefont {Salton}\ \emph {et~al.}(2015)\citenamefont {Salton},
  \citenamefont {Mann},\ and\ \citenamefont {Menicucci}}]{Salton:2014jaa}%
  \BibitemOpen
  \bibfield  {author} {\bibinfo {author} {\bibfnamefont {G.}~\bibnamefont
  {Salton}}, \bibinfo {author} {\bibfnamefont {R.~B.}\ \bibnamefont {Mann}},\
  and\ \bibinfo {author} {\bibfnamefont {N.~C.}\ \bibnamefont {Menicucci}},\
  }\bibfield  {title} {\bibinfo {title} {{Acceleration-assisted entanglement
  harvesting and rangefinding}},\ }\href
  {https://doi.org/10.1088/1367-2630/17/3/035001} {\bibfield  {journal}
  {\bibinfo  {journal} {New J. Phys.}\ }\textbf {\bibinfo {volume} {17}},\
  \bibinfo {pages} {035001} (\bibinfo {year} {2015})}\BibitemShut {NoStop}%
\bibitem [{\citenamefont {Pozas-Kerstjens}\ and\ \citenamefont
  {Mart\'{i}n-Mart\'{i}nez}(2016)}]{Pozas2016}%
  \BibitemOpen
  \bibfield  {author} {\bibinfo {author} {\bibfnamefont {A.}~\bibnamefont
  {Pozas-Kerstjens}}\ and\ \bibinfo {author} {\bibfnamefont {E.}~\bibnamefont
  {Mart\'{i}n-Mart\'{i}nez}},\ }\bibfield  {title} {\bibinfo {title}
  {Entanglement harvesting from the electromagnetic vacuum with hydrogenlike
  atoms},\ }\href {https://doi.org/10.1103/PhysRevD.94.064074} {\bibfield
  {journal} {\bibinfo  {journal} {Phys. Rev. D}\ }\textbf {\bibinfo {volume}
  {94}},\ \bibinfo {pages} {064074} (\bibinfo {year} {2016})}\BibitemShut
  {NoStop}%
\bibitem [{\citenamefont {Simidzija}\ and\ \citenamefont
  {Mart\'{\i}n-Mart\'{\i}nez}(2017)}]{PetarOld}%
  \BibitemOpen
  \bibfield  {author} {\bibinfo {author} {\bibfnamefont {P.}~\bibnamefont
  {Simidzija}}\ and\ \bibinfo {author} {\bibfnamefont {E.}~\bibnamefont
  {Mart\'{\i}n-Mart\'{\i}nez}},\ }\bibfield  {title} {\bibinfo {title} {All
  coherent field states entangle equally},\ }\href
  {https://doi.org/10.1103/PhysRevD.96.025020} {\bibfield  {journal} {\bibinfo
  {journal} {Phys. Rev. D}\ }\textbf {\bibinfo {volume} {96}},\ \bibinfo
  {pages} {025020} (\bibinfo {year} {2017})}\BibitemShut {NoStop}%
\bibitem [{\citenamefont {Simidzija}\ and\ \citenamefont
  {Mart\'{i}n-Mart\'{i}nez}(2018)}]{Petar}%
  \BibitemOpen
  \bibfield  {author} {\bibinfo {author} {\bibfnamefont {P.}~\bibnamefont
  {Simidzija}}\ and\ \bibinfo {author} {\bibfnamefont {E.}~\bibnamefont
  {Mart\'{i}n-Mart\'{i}nez}},\ }\bibfield  {title} {\bibinfo {title}
  {Harvesting correlations from thermal and squeezed coherent states},\ }\href
  {https://doi.org/10.1103/PhysRevD.98.085007} {\bibfield  {journal} {\bibinfo
  {journal} {Phys. Rev. D}\ }\textbf {\bibinfo {volume} {98}},\ \bibinfo
  {pages} {085007} (\bibinfo {year} {2018})}\BibitemShut {NoStop}%
\bibitem [{\citenamefont {Henderson}\ \emph {et~al.}(2018)\citenamefont
  {Henderson}, \citenamefont {Hennigar}, \citenamefont {Mann}, \citenamefont
  {Smith},\ and\ \citenamefont {Zhang}}]{Henderson2018}%
  \BibitemOpen
  \bibfield  {author} {\bibinfo {author} {\bibfnamefont {L.~J.}\ \bibnamefont
  {Henderson}}, \bibinfo {author} {\bibfnamefont {R.~A.}\ \bibnamefont
  {Hennigar}}, \bibinfo {author} {\bibfnamefont {R.~B.}\ \bibnamefont {Mann}},
  \bibinfo {author} {\bibfnamefont {A.~R.~H.}\ \bibnamefont {Smith}},\ and\
  \bibinfo {author} {\bibfnamefont {J.}~\bibnamefont {Zhang}},\ }\bibfield
  {title} {\bibinfo {title} {Harvesting entanglement from the black hole
  vacuum},\ }\href {https://doi.org/10.1088/1361-6382/aae27e} {\bibfield
  {journal} {\bibinfo  {journal} {Class. Quantum Gravity}\ }\textbf {\bibinfo
  {volume} {35}},\ \bibinfo {pages} {21LT02} (\bibinfo {year}
  {2018})}\BibitemShut {NoStop}%
\bibitem [{\citenamefont {Ng}\ \emph {et~al.}(2018{\natexlab{a}})\citenamefont
  {Ng}, \citenamefont {Mann},\ and\ \citenamefont
  {Mart\'{\i}n-Mart\'{\i}nez}}]{Ng2018}%
  \BibitemOpen
  \bibfield  {author} {\bibinfo {author} {\bibfnamefont {K.~K.}\ \bibnamefont
  {Ng}}, \bibinfo {author} {\bibfnamefont {R.~B.}\ \bibnamefont {Mann}},\ and\
  \bibinfo {author} {\bibfnamefont {E.}~\bibnamefont
  {Mart\'{\i}n-Mart\'{\i}nez}},\ }\bibfield  {title} {\bibinfo {title} {New
  techniques for entanglement harvesting in flat and curved spacetimes},\
  }\href {https://doi.org/10.1103/PhysRevD.97.125011} {\bibfield  {journal}
  {\bibinfo  {journal} {Phys. Rev. D}\ }\textbf {\bibinfo {volume} {97}},\
  \bibinfo {pages} {125011} (\bibinfo {year} {2018}{\natexlab{a}})}\BibitemShut
  {NoStop}%
\bibitem [{\citenamefont {Ng}\ \emph {et~al.}(2018{\natexlab{b}})\citenamefont
  {Ng}, \citenamefont {Mann},\ and\ \citenamefont
  {Mart\'{i}n-Mart\'{i}nez}}]{Ng2}%
  \BibitemOpen
  \bibfield  {author} {\bibinfo {author} {\bibfnamefont {K.~K.}\ \bibnamefont
  {Ng}}, \bibinfo {author} {\bibfnamefont {R.~B.}\ \bibnamefont {Mann}},\ and\
  \bibinfo {author} {\bibfnamefont {E.}~\bibnamefont
  {Mart\'{i}n-Mart\'{i}nez}},\ }\bibfield  {title} {\bibinfo {title}
  {Unruh-{D}e{W}itt detectors and entanglement: The anti--de {S}itter space},\
  }\href {https://doi.org/10.1103/PhysRevD.98.125005} {\bibfield  {journal}
  {\bibinfo  {journal} {Phys. Rev. D}\ }\textbf {\bibinfo {volume} {98}},\
  \bibinfo {pages} {125005} (\bibinfo {year} {2018}{\natexlab{b}})}\BibitemShut
  {NoStop}%
\bibitem [{\citenamefont {Henderson}\ \emph {et~al.}(2019)\citenamefont
  {Henderson}, \citenamefont {Hennigar}, \citenamefont {Mann}, \citenamefont
  {Smith},\ and\ \citenamefont {Zhang}}]{Henderson2019}%
  \BibitemOpen
  \bibfield  {author} {\bibinfo {author} {\bibfnamefont {L.~J.}\ \bibnamefont
  {Henderson}}, \bibinfo {author} {\bibfnamefont {R.~A.}\ \bibnamefont
  {Hennigar}}, \bibinfo {author} {\bibfnamefont {R.~B.}\ \bibnamefont {Mann}},
  \bibinfo {author} {\bibfnamefont {A.~R.~H.}\ \bibnamefont {Smith}},\ and\
  \bibinfo {author} {\bibfnamefont {J.}~\bibnamefont {Zhang}},\ }\bibfield
  {title} {\bibinfo {title} {Entangling detectors in anti-de {S}itter space},\
  }\href {https://doi.org/10.1007/JHEP05(2019)178} {\bibfield  {journal}
  {\bibinfo  {journal} {J. High Energy Phys.}\ }\textbf {\bibinfo {volume}
  {2019}},\ \bibinfo {pages} {178}}\BibitemShut {NoStop}%
\bibitem [{\citenamefont {Cong}\ \emph {et~al.}(2019)\citenamefont {Cong},
  \citenamefont {Tjoa},\ and\ \citenamefont {Mann}}]{Cong2019}%
  \BibitemOpen
  \bibfield  {author} {\bibinfo {author} {\bibfnamefont {W.}~\bibnamefont
  {Cong}}, \bibinfo {author} {\bibfnamefont {E.}~\bibnamefont {Tjoa}},\ and\
  \bibinfo {author} {\bibfnamefont {R.~B.}\ \bibnamefont {Mann}},\ }\bibfield
  {title} {\bibinfo {title} {Entanglement harvesting with moving mirrors},\
  }\href {https://doi.org/10.1007\%2Fjhep06\%282019\%29021} {\bibfield
  {journal} {\bibinfo  {journal} {J. High Energy Phys.}\ }\textbf {\bibinfo
  {volume} {2019}},\ \bibinfo {pages} {21}}\BibitemShut {NoStop}%
\bibitem [{\citenamefont {Cong}\ \emph {et~al.}(2020)\citenamefont {Cong},
  \citenamefont {Qian}, \citenamefont {Good},\ and\ \citenamefont
  {Mann}}]{Cong2020}%
  \BibitemOpen
  \bibfield  {author} {\bibinfo {author} {\bibfnamefont {W.}~\bibnamefont
  {Cong}}, \bibinfo {author} {\bibfnamefont {C.}~\bibnamefont {Qian}}, \bibinfo
  {author} {\bibfnamefont {M.~R.}\ \bibnamefont {Good}},\ and\ \bibinfo
  {author} {\bibfnamefont {R.~B.}\ \bibnamefont {Mann}},\ }\bibfield  {title}
  {\bibinfo {title} {Effects of horizons on entanglement harvesting},\ }\href
  {https://doi.org/10.1007\%2Fjhep10\%282020\%29067} {\bibfield  {journal}
  {\bibinfo  {journal} {J. High Energy Phys.}\ }\textbf {\bibinfo {volume}
  {2020}},\ \bibinfo {pages} {67}}\BibitemShut {NoStop}%
\bibitem [{\citenamefont {Henderson}\ and\ \citenamefont
  {Menicucci}(2020)}]{Henderson2020}%
  \BibitemOpen
  \bibfield  {author} {\bibinfo {author} {\bibfnamefont {L.~J.}\ \bibnamefont
  {Henderson}}\ and\ \bibinfo {author} {\bibfnamefont {N.~C.}\ \bibnamefont
  {Menicucci}},\ }\bibfield  {title} {\bibinfo {title} {Bandlimited
  entanglement harvesting},\ }\href
  {https://doi.org/10.1103/PhysRevD.102.125026} {\bibfield  {journal} {\bibinfo
   {journal} {Phys. Rev. D}\ }\textbf {\bibinfo {volume} {102}},\ \bibinfo
  {pages} {125026} (\bibinfo {year} {2020})}\BibitemShut {NoStop}%
\bibitem [{\citenamefont {Tjoa}\ and\ \citenamefont
  {Mart\'{\i}n-Mart\'{\i}nez}(2021)}]{ericksonNew}%
  \BibitemOpen
  \bibfield  {author} {\bibinfo {author} {\bibfnamefont {E.}~\bibnamefont
  {Tjoa}}\ and\ \bibinfo {author} {\bibfnamefont {E.}~\bibnamefont
  {Mart\'{\i}n-Mart\'{\i}nez}},\ }\bibfield  {title} {\bibinfo {title} {When
  entanglement harvesting is not really harvesting},\ }\href
  {https://doi.org/10.1103/PhysRevD.104.125005} {\bibfield  {journal} {\bibinfo
   {journal} {Phys. Rev. D}\ }\textbf {\bibinfo {volume} {104}},\ \bibinfo
  {pages} {125005} (\bibinfo {year} {2021})}\BibitemShut {NoStop}%
\bibitem [{\citenamefont {Liu}\ \emph {et~al.}(2021)\citenamefont {Liu},
  \citenamefont {Zhang},\ and\ \citenamefont {Yu}}]{Liu2021}%
  \BibitemOpen
  \bibfield  {author} {\bibinfo {author} {\bibfnamefont {Z.}~\bibnamefont
  {Liu}}, \bibinfo {author} {\bibfnamefont {J.}~\bibnamefont {Zhang}},\ and\
  \bibinfo {author} {\bibfnamefont {H.}~\bibnamefont {Yu}},\ }\bibfield
  {title} {\bibinfo {title} {Entanglement harvesting in the presence of a
  reflecting boundary},\ }\href
  {https://doi.org/10.1007\%2Fjhep08\%282021\%29020} {\bibfield  {journal}
  {\bibinfo  {journal} {J. High Energy Phys.}\ }\textbf {\bibinfo {volume}
  {2021}},\ \bibinfo {pages} {20}}\BibitemShut {NoStop}%
\bibitem [{\citenamefont {Perche}\ \emph {et~al.}(2022)\citenamefont {Perche},
  \citenamefont {Lima},\ and\ \citenamefont
  {Mart\'{\i}n-Mart\'{\i}nez}}]{carol}%
  \BibitemOpen
  \bibfield  {author} {\bibinfo {author} {\bibfnamefont {T.~R.}\ \bibnamefont
  {Perche}}, \bibinfo {author} {\bibfnamefont {C.}~\bibnamefont {Lima}},\ and\
  \bibinfo {author} {\bibfnamefont {E.}~\bibnamefont
  {Mart\'{\i}n-Mart\'{\i}nez}},\ }\bibfield  {title} {\bibinfo {title}
  {Harvesting entanglement from complex scalar and fermionic fields with
  linearly coupled particle detectors},\ }\href
  {https://doi.org/10.1103/PhysRevD.105.065016} {\bibfield  {journal} {\bibinfo
   {journal} {Phys. Rev. D}\ }\textbf {\bibinfo {volume} {105}},\ \bibinfo
  {pages} {065016} (\bibinfo {year} {2022})}\BibitemShut {NoStop}%
\bibitem [{\citenamefont {Mart\'{\i}n-Mart\'{\i}nez}\ \emph
  {et~al.}(2013)\citenamefont {Mart\'{\i}n-Mart\'{\i}nez}, \citenamefont
  {Brown}, \citenamefont {Donnelly},\ and\ \citenamefont
  {Kempf}}]{Edu2013farming}%
  \BibitemOpen
  \bibfield  {author} {\bibinfo {author} {\bibfnamefont {E.}~\bibnamefont
  {Mart\'{\i}n-Mart\'{\i}nez}}, \bibinfo {author} {\bibfnamefont {E.~G.}\
  \bibnamefont {Brown}}, \bibinfo {author} {\bibfnamefont {W.}~\bibnamefont
  {Donnelly}},\ and\ \bibinfo {author} {\bibfnamefont {A.}~\bibnamefont
  {Kempf}},\ }\bibfield  {title} {\bibinfo {title} {Sustainable entanglement
  production from a quantum field},\ }\href
  {https://doi.org/10.1103/PhysRevA.88.052310} {\bibfield  {journal} {\bibinfo
  {journal} {Phys. Rev. A}\ }\textbf {\bibinfo {volume} {88}},\ \bibinfo
  {pages} {052310} (\bibinfo {year} {2013})}\BibitemShut {NoStop}%
\bibitem [{\citenamefont {Brown}\ \emph {et~al.}(2014)\citenamefont {Brown},
  \citenamefont {Donnelly}, \citenamefont {Kempf}, \citenamefont {Mann},
  \citenamefont {Mart{\'{\i}}n-Mart{\'{\i}}nez},\ and\ \citenamefont
  {Menicucci}}]{Brown2014}%
  \BibitemOpen
  \bibfield  {author} {\bibinfo {author} {\bibfnamefont {E.~G.}\ \bibnamefont
  {Brown}}, \bibinfo {author} {\bibfnamefont {W.}~\bibnamefont {Donnelly}},
  \bibinfo {author} {\bibfnamefont {A.}~\bibnamefont {Kempf}}, \bibinfo
  {author} {\bibfnamefont {R.~B.}\ \bibnamefont {Mann}}, \bibinfo {author}
  {\bibfnamefont {E.}~\bibnamefont {Mart{\'{\i}}n-Mart{\'{\i}}nez}},\ and\
  \bibinfo {author} {\bibfnamefont {N.~C.}\ \bibnamefont {Menicucci}},\
  }\bibfield  {title} {\bibinfo {title} {Quantum seismology},\ }\href
  {https://doi.org/10.1088/1367-2630/16/10/105020} {\bibfield  {journal}
  {\bibinfo  {journal} {New Journal of Physics}\ }\textbf {\bibinfo {volume}
  {16}},\ \bibinfo {pages} {105020} (\bibinfo {year} {2014})}\BibitemShut
  {NoStop}%
\bibitem [{\citenamefont {Mart\'{\i}n-Mart\'{\i}nez}\ \emph
  {et~al.}(2016)\citenamefont {Mart\'{\i}n-Mart\'{\i}nez}, \citenamefont
  {Smith},\ and\ \citenamefont {Terno}}]{topology}%
  \BibitemOpen
  \bibfield  {author} {\bibinfo {author} {\bibfnamefont {E.}~\bibnamefont
  {Mart\'{\i}n-Mart\'{\i}nez}}, \bibinfo {author} {\bibfnamefont {A.~R.~H.}\
  \bibnamefont {Smith}},\ and\ \bibinfo {author} {\bibfnamefont {D.~R.}\
  \bibnamefont {Terno}},\ }\bibfield  {title} {\bibinfo {title} {Spacetime
  structure and vacuum entanglement},\ }\href
  {https://doi.org/10.1103/PhysRevD.93.044001} {\bibfield  {journal} {\bibinfo
  {journal} {Phys. Rev. D}\ }\textbf {\bibinfo {volume} {93}},\ \bibinfo
  {pages} {044001} (\bibinfo {year} {2016})}\BibitemShut {NoStop}%
\bibitem [{\citenamefont
  {Mart\'{i}n-Mart\'{i}nez}(2015)}]{martin-martinez2015}%
  \BibitemOpen
  \bibfield  {author} {\bibinfo {author} {\bibfnamefont {E.}~\bibnamefont
  {Mart\'{i}n-Mart\'{i}nez}},\ }\bibfield  {title} {\bibinfo {title} {Causality
  issues of particle detector models in {QFT} and quantum optics},\ }\href
  {https://doi.org/10.1103/PhysRevD.92.104019} {\bibfield  {journal} {\bibinfo
  {journal} {Phys. Rev. D}\ }\textbf {\bibinfo {volume} {92}},\ \bibinfo
  {pages} {104019} (\bibinfo {year} {2015})}\BibitemShut {NoStop}%
\bibitem [{\citenamefont {de~Ram\'on}\ \emph {et~al.}(2021)\citenamefont
  {de~Ram\'on}, \citenamefont {Papageorgiou},\ and\ \citenamefont
  {Mart\'{\i}n-Mart\'{\i}nez}}]{pipoFTL}%
  \BibitemOpen
  \bibfield  {author} {\bibinfo {author} {\bibfnamefont {J.}~\bibnamefont
  {de~Ram\'on}}, \bibinfo {author} {\bibfnamefont {M.}~\bibnamefont
  {Papageorgiou}},\ and\ \bibinfo {author} {\bibfnamefont {E.}~\bibnamefont
  {Mart\'{\i}n-Mart\'{\i}nez}},\ }\bibfield  {title} {\bibinfo {title}
  {Relativistic causality in particle detector models: Faster-than-light
  signaling and impossible measurements},\ }\href
  {https://doi.org/10.1103/PhysRevD.103.085002} {\bibfield  {journal} {\bibinfo
   {journal} {Phys. Rev. D}\ }\textbf {\bibinfo {volume} {103}},\ \bibinfo
  {pages} {085002} (\bibinfo {year} {2021})}\BibitemShut {NoStop}%
\bibitem [{\citenamefont {Mart\'{i}n-Mart\'{i}nez}\ \emph
  {et~al.}(2020)\citenamefont {Mart\'{i}n-Mart\'{i}nez}, \citenamefont
  {Perche},\ and\ \citenamefont {de~S.~L.~Torres}}]{us}%
  \BibitemOpen
  \bibfield  {author} {\bibinfo {author} {\bibfnamefont {E.}~\bibnamefont
  {Mart\'{i}n-Mart\'{i}nez}}, \bibinfo {author} {\bibfnamefont {T.~R.}\
  \bibnamefont {Perche}},\ and\ \bibinfo {author} {\bibfnamefont
  {B.}~\bibnamefont {de~S.~L.~Torres}},\ }\bibfield  {title} {\bibinfo {title}
  {General relativistic quantum optics: Finite-size particle detector models in
  curved spacetimes},\ }\href {https://doi.org/10.1103/PhysRevD.101.045017}
  {\bibfield  {journal} {\bibinfo  {journal} {Phys. Rev. D}\ }\textbf {\bibinfo
  {volume} {101}},\ \bibinfo {pages} {045017} (\bibinfo {year}
  {2020})}\BibitemShut {NoStop}%
\bibitem [{\citenamefont {Mart\'{\i}n-Mart\'{\i}nez}\ \emph
  {et~al.}(2021)\citenamefont {Mart\'{\i}n-Mart\'{\i}nez}, \citenamefont
  {Perche},\ and\ \citenamefont {Torres}}]{us2}%
  \BibitemOpen
  \bibfield  {author} {\bibinfo {author} {\bibfnamefont {E.}~\bibnamefont
  {Mart\'{\i}n-Mart\'{\i}nez}}, \bibinfo {author} {\bibfnamefont {T.~R.}\
  \bibnamefont {Perche}},\ and\ \bibinfo {author} {\bibfnamefont {B.~d. S.~L.}\
  \bibnamefont {Torres}},\ }\bibfield  {title} {\bibinfo {title} {Broken
  covariance of particle detector models in relativistic quantum information},\
  }\href {https://doi.org/10.1103/PhysRevD.103.025007} {\bibfield  {journal}
  {\bibinfo  {journal} {Phys. Rev. D}\ }\textbf {\bibinfo {volume} {103}},\
  \bibinfo {pages} {025007} (\bibinfo {year} {2021})}\BibitemShut {NoStop}%
\bibitem [{\citenamefont {Fulling}(1989)}]{Fulling1989}%
  \BibitemOpen
  \bibfield  {author} {\bibinfo {author} {\bibfnamefont {S.~A.}\ \bibnamefont
  {Fulling}},\ }\href@noop {} {\emph {\bibinfo {title} {Aspects of Quantum
  Field Theory in Curved Spacetime}}},\ London Mathematical Society Student
  Texts\ (\bibinfo  {publisher} {Cambridge University Press},\ \bibinfo {year}
  {1989})\BibitemShut {NoStop}%
\bibitem [{\citenamefont {Wald}(1994)}]{Wald2}%
  \BibitemOpen
  \bibfield  {author} {\bibinfo {author} {\bibfnamefont {R.~M.}\ \bibnamefont
  {Wald}},\ }\href@noop {} {\emph {\bibinfo {title} {Quantum Field Theory in
  Curved Spacetime and Black Hole Thermodynamics}}}\ (\bibinfo  {publisher}
  {The University of Chicago Press},\ \bibinfo {year} {1994})\BibitemShut
  {NoStop}%
\bibitem [{\citenamefont {Vidal}\ and\ \citenamefont
  {Werner}(2002)}]{Vidal2002}%
  \BibitemOpen
  \bibfield  {author} {\bibinfo {author} {\bibfnamefont {G.}~\bibnamefont
  {Vidal}}\ and\ \bibinfo {author} {\bibfnamefont {R.~F.}\ \bibnamefont
  {Werner}},\ }\bibfield  {title} {\bibinfo {title} {Computable measure of
  entanglement},\ }\href {https://doi.org/10.1103/PhysRevA.65.032314}
  {\bibfield  {journal} {\bibinfo  {journal} {Phys. Rev. A}\ }\textbf {\bibinfo
  {volume} {65}},\ \bibinfo {pages} {032314} (\bibinfo {year}
  {2002})}\BibitemShut {NoStop}%
\bibitem [{\citenamefont {Grimmer}\ \emph {et~al.}(2021)\citenamefont
  {Grimmer}, \citenamefont {Torres},\ and\ \citenamefont
  {Mart\'{\i}n-Mart\'{\i}nez}}]{BrunoDan}%
  \BibitemOpen
  \bibfield  {author} {\bibinfo {author} {\bibfnamefont {D.}~\bibnamefont
  {Grimmer}}, \bibinfo {author} {\bibfnamefont {B.~d. S.~L.}\ \bibnamefont
  {Torres}},\ and\ \bibinfo {author} {\bibfnamefont {E.}~\bibnamefont
  {Mart\'{\i}n-Mart\'{\i}nez}},\ }\bibfield  {title} {\bibinfo {title}
  {Measurements in {QFT}: {W}eakly coupled local particle detectors and
  entanglement harvesting},\ }\href
  {https://doi.org/10.1103/PhysRevD.104.085014} {\bibfield  {journal} {\bibinfo
   {journal} {Phys. Rev. D}\ }\textbf {\bibinfo {volume} {104}},\ \bibinfo
  {pages} {085014} (\bibinfo {year} {2021})}\BibitemShut {NoStop}%
\end{thebibliography}%
    
\end{document}